\documentclass[sigconf]{acmart}

\usepackage{color}
\usepackage{amsmath}
\usepackage{tcolorbox}
\usepackage{multirow}
\usepackage{multicol}
\usepackage{amsmath,amssymb,amsfonts}
\usepackage{graphicx}
\usepackage{textcomp}
\usepackage{booktabs}
\usepackage[caption=false]{subfig}
\usepackage{url}
\usepackage{balance}
\usepackage{xspace}
\usepackage{float}

\newcommand{\eg}{\emph{e.g.,}\xspace}
\newcommand{\ie}{\emph{i.e.,}\xspace}

\usepackage{changepage}
\usepackage{framed}
\makeatletter
{\par\unskip\endMakeFramed}
\makeatother
\definecolor{formalshade}{rgb}{0.93,0.93,0.93}
\definecolor{darkblue}{rgb}{0.2, 0.2, 0.2}
\newenvironment{formal}{%
\def\FrameCommand{%
  \hspace{1pt}%
  {\color{darkblue}\vrule width 2pt}%
  {\color{formalshade}\vrule width 4pt}%
  \colorbox{formalshade}%
}%
\MakeFramed{\advance\hsize-\width\FrameRestore}%
\noindent\hspace{-1pt}
\begin{adjustwidth}{}{7pt}%
\vspace{2pt}\vspace{2pt}%
}
{%
\vspace{3pt}\end{adjustwidth}\endMakeFramed%
}
\newcounter{resultcounter}

\newcounter{patterncounter}

\copyrightyear{2020} 
\acmYear{2020} 
\setcopyright{acmcopyright}
\acmConference[MSR '20]{17th International Conference on Mining Software Repositories}{October 5--6, 2020}{Seoul, Republic of Korea}
\acmBooktitle{17th International Conference on Mining Software Repositories (MSR '20), October 5--6, 2020, Seoul, Republic of Korea}
\acmPrice{15.00}
\acmDOI{10.1145/3379597.3387459}
\acmISBN{978-1-4503-7517-7/20/05}


\begin{document}


\title{Beyond the Code: Mining Self-Admitted Technical \\Debt in Issue Tracker Systems}


\author{Laerte Xavier}
\affiliation{
ASERG Group - Department of Computer Science\\
Federal University of Minas Gerais (UFMG)\\
Belo Horizonte, Brazil}
\email{laertexavier@dcc.ufmg.br}

\author{Fabio Ferreira}
\affiliation{
Center of Informatics - Federal Institute\\
of the Southeast of Minas Gerais\\
Barbacena, Brazil}
\email{fabio.ferreira@ifsudestemg.edu.br}

\author{Rodrigo Brito}
\affiliation{
ASERG Group - Department of Computer Science\\
Federal University of Minas Gerais (UFMG)\\
Belo Horizonte, Brazil}
\email{britorodrigo@dcc.ufmg.br}

\author{Marco Tulio Valente}
\affiliation{
ASERG Group - Department of Computer Science\\
Federal University of Minas Gerais (UFMG)\\
Belo Horizonte, Brazil}
\email{mtov@dcc.ufmg.br}


\begin{abstract}

Self-admitted technical debt (SATD) is a particular case of Technical Debt (TD) where 
developers explicitly acknowledge their sub-optimal implementation decisions.
Previous studies mine SATD by searching for specific TD-related terms in source 
code comments.
By contrast, in this paper we argue that developers can admit technical 
debt by other means, e.g., by creating issues in tracking systems
and labelling them as referring to TD. We refer to this type of SATD as
issue-based SATD or just SATD-I. We study a sample of 286
SATD-I instances collected from five open source projects,
including Microsoft Visual Studio and GitLab Community Edition.
We show that only 29\% of the studied SATD-I instances can be
tracked to source code comments. We also show that SATD-I issues take more time to be closed, compared to other issues, although they are not more complex in terms of code churn. Besides, in 45\% of the studied issues TD was introduced to ship earlier, and in almost 60\% it refers to {\sc Design} flaws. Finally, we report that most developers pay SATD-I to reduce its costs or interests (66\%).
Our findings  suggest that there is space for designing  novel tools to support technical debt management, particularly tools that encourage developers to create and label issues containing TD concerns.

\end{abstract}


\maketitle

\section{Introduction}
\label{sec:intro}

In software development, the ``done is better than perfect'' maxim reflects the inevitable trade-off between keeping software quality and releasing on time.
In this context, the Technical Debt (TD) metaphor---first framed by Cunningham in 1992~\cite{td-first}---refers to the unavoidable maintenance and evolution costs of such {\it not-quite-right} solutions.
Several circumstances drive developers to assume these debts, such as deadline pressure, existing low quality code, and poor software process~\cite{balancing-act}. 
In fact, the term has been widely adopted since its definition~\cite{td-definitions} and became subject of various studies, mostly regarding its identification~\cite{td-case-study, rw14, rw21, td-id:ist}, management~\cite{rw10, todo-tobug, rw16}, and impact~\cite{td-impact, rw8, td-analytics, msr2016-bellomo, td-pr}.

\begin{figure*}[t]
	\centering
    \includegraphics[width=0.938\textwidth]{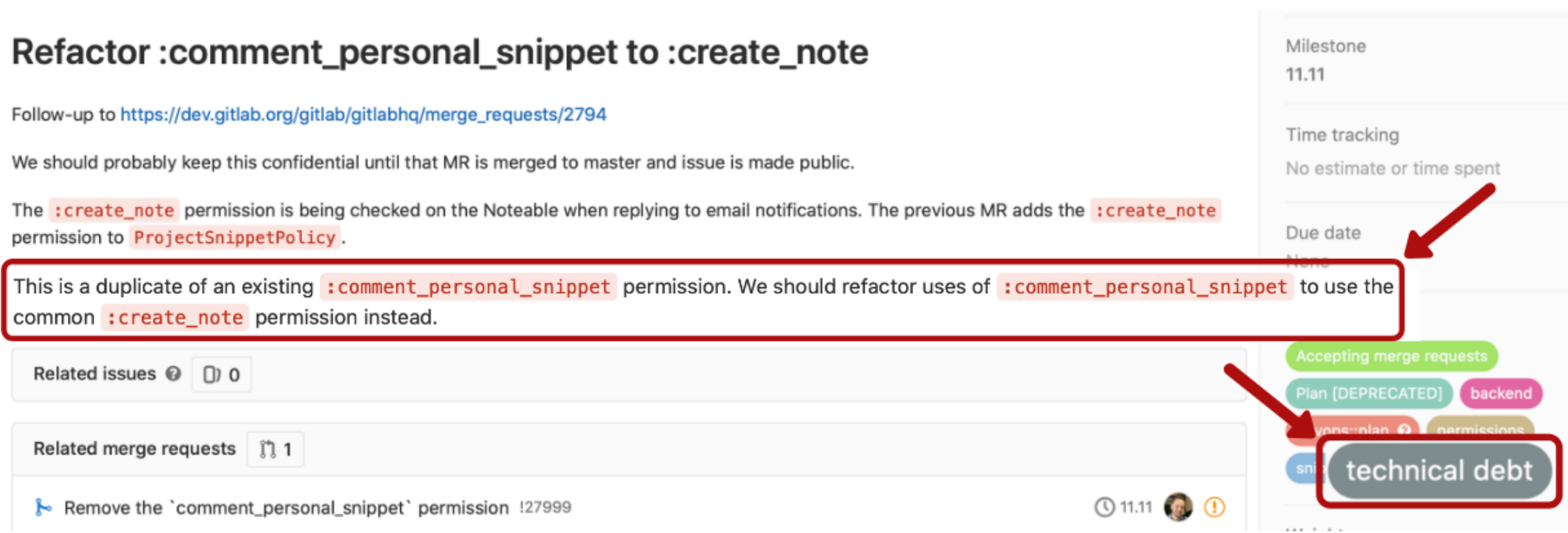}
	\caption{Example of SATD in a GitLab's issue}
	\label{fig:td-issues}
	\vspace{-3mm}
\end{figure*}

Self-admitted technical debt (SATD) is a particular case of TD where developers explicitly admit their sub-optimal implementation decisions~\cite{potdar2014, rw5, rw20, rw2}. However, to our knowledge, SATD studies rely only on source comments to identify SATD instances. Particularly, existing studies search for specific TD-related terms in source code comments --- such as {\em fixme}, {\em TODO}, and {\em to be fixed}. By contrast, in this paper we argue that developers can acknowledge technical debt out of the source code, by creating issues in tracking systems documenting their sub-optimal implementation decisions. To document the debt, developers label these issues with terms such as {\em technical debt} or {\em debt}. An example is presented in Figure~\ref{fig:td-issues}, in the next page. The figure shows an issue from GitLab requesting the removal of duplicated code (in this case, a permission variable). 
As we can see, it received a Technical Debt label.

We assume there are at least two types of SATD:
(1) {\bf SATD documented using source code comments (SATD-C)}, which has been extensively studied in the past; and (2) {\bf SATD documented using issues (STAD-I)}, which we propose to study in this paper. Moreover, we focus our study on the payment of SATD-I, i.e., only on issues documenting TD that was successfully closed by developers, indicating the admitted TD problem was solved. With this decision, our intention is to study  SATD-I instances that had a practical and positive impact on the projects. It also allowed us to investigate the overlap between SATD-I and SATD-C.


We collect and characterize 286 SATD-I instances from five relevant open-source systems, including GitLab (a git-hosting platform that is publicly developed and maintained using its own services), and VS Code (the popular IDE from Microsoft). Developers of these repositories follow a practice to create and label issues that refer to TD problems, i.e., we view these instances as cases of SATD-I. We use this dataset to answer four research questions:

\vspace{1.5mm}
\noindent\textbf{RQ1. What is the overlap between SATD-C and SATD-I?}\\ Our intention is to check whether SATD-C and SATD-I refer to different cases of TD. To this purpose, we check whether SATD-I instances are also documented in the source code, i.e., whether there is a correspondence from SATD-I instances to SATD-C found in the source code of the studied systems. To identify this correspondence, we relied on a state-of-the-art tool to detect SATD-C.\vspace{1mm}

\noindent\textbf{RQ2. What types of technical debt are paid in SATD-I?}\\ In this RQ, we manually analyze and classify the TD problems
documented, discussed, and fixed in 
the 286 instances of SATD-I from our dataset.
To perform this classification, we reuse 
ten categories of TD from the literature~\cite{slr-td}.\vspace{1mm}

\noindent\textbf{RQ3. Why do developers introduce SATD-I?}\\ Next, we perform a survey with developers directly involved on SATD-I payment. We analyze 30 received answers (response rate of $\sim$35\%) describing why they introduced the studied SATD-I.\vspace{1mm}

\noindent\textbf{RQ4. Why do developers pay SATD-I?}\\ We also elicit a list of five main reasons that drive developers to pay SATD-I by asking the participants of our survey why they decided to close the studied issues. We also shed light on TD interests by investigating the maintenance problems caused by SATD-I.\vspace{1mm}


We make three key contributions in this paper:

\begin{itemize}

\item We identify and study SATD beyond the source code, i.e., detected by mining issue tracker systems. We confirm that developers use issues to admit technical debt in their projects, i.e., SATD does not appear only in code.
Finally, we show that there is an overlap between SATD-I and SATD-C but it is not dominant. Only 29\% of the studied SATD-I instances can be  traced to SATD-C instances.      
    
\item We show that SATD-I instances take more time to be closed, compared to other issues, although they are not more complex in terms of code churn. Besides, TD was deliberately introduced in 45\% of the studied SATD-I instances, and in almost 60\% of them it refers to {\sc Design} flaws. We also found that most developers paid SATD-I to reduce its interests (66\%), and to have a clean code (28\%). 

\item Our findings  suggest that there is space for designing and implementing novel tools to support technical debt management, particularly tools that encourage developers to create and label issues containing TD concerns, i.e., to self-admit TD using issues. Moreover, our findings reinforce the importance of introducing TD payment activities as part of software development processes, as a strategy to preserve internal quality or to introduce newcomers to the codebase.
\end{itemize}

\noindent\textit{Structure of the paper.}
In Section~\ref{sec:definition} we detail our definition of SATD-I.
Section~\ref{sec:data} presents our dataset of SATD-I instances.
In Sections~\ref{sec:satd},~\ref{sec:class} and~\ref{sec:quali}, we answer the proposed research questions.
Section~\ref{sec:discussion} provides the main implications of our findings.
In Sections~\ref{sec:threats} and~\ref{sec:related-work}, we describe threats to validity and related work, respectively.
Finally, Section~\ref{sec:conclusion} concludes our work.
\vspace{-2mm}

\section{Definition}
\label{sec:definition}

We define SATD-I as technical debt instances documented using issues.
It contrasts with SATD-C, which is documented using source code comments.
Particularly, this definition does not require the developer who introduced the debt in the source to be the same developer responsible for creating the issue in the tracking system. A similar rule is followed by studies targeting SATD-C, which do not check whether the TD-related code and the TD-related comments were inserted by the same author~\cite{potdar2014, rw20,rw2}. Therefore, SATD should be viewed as
TD acknowledged and detected by the own developers of a software project, without the help of any tool.
SATD contrasts, for example, with TD automatically detected using 
static analysis tools~\cite{td-case-study, rw10}.
\vspace{-2mm}

\section{Dataset}
\label{sec:data}

In this paper, we study SATD-I instances from five open-source systems: GitLab and four GitHub-based systems.
We selected GitLab because it is a well-known platform that supports a git-based version control service and also a CI/CD pipeline. Moreover, we had previous knowledge---from our research in the area---on GitLab's practice to label TD-related issues. 

In this section, we explain how we selected the GitLab issues used in this study (Section~\ref{sec:gitlab}). We also explain how we mined and selected four GitHub projects that follow a practice similar to the one used by GitLab, i.e., they also use specific labels on issues that discuss technical debt (Section~\ref{sec:github}). Finally, we provide a quantitative overview of the proposed dataset (Section~\ref{sec:characterization}).

\subsection{GitLab CE} 
\label{sec:gitlab}

Differently from GitHub, GitLab's source code is publicly available in the platform, i.e., GitLab is an open source project that is developed and maintained using its own services. 
In fact, the project has two editions: Community (CE) and Enterprise (EE).  
The latter is a commercial version and the former is  an open-source edition.
GitLab's development happens on both repositories, which are continuously synchronized. 
Since they are public, we rely on  issues from GitLab CE. 

First, we used GitLab's REST API to select all issues with a {\em technical debt} label that were {\em closed} in the last {\em six months} (December 15th, 2018 to June 15th, 2019). We only selected closed issues because our focus is on technical debt that was paid. Moreover, we restricted the selection to the last six months to increase the chances of receiving answers in the survey we performed with GitLab's developers---and also to increase the confidence on the survey answers (see Section~\ref{sec:quali}).

After applying the described selection criteria, we found 188 issues. 
The first author of this paper carefully inspected each one and removed 65 issues (34.6\%) that represent duplicated issues, issues that only include discussions, and ignored issues. He also verified that no issue was automatically tagged by a static analysis tool. For example, he discarded an issue where the developer concluded that:

\vspace{0.6mm}
\noindent{\em Heh, this is a duplicate of gitlab-ee\#3861 (closed), which is being worked on right now by @cablett. I'll close it!}\footnote{\url{https://gitlab.com/gitlab-org/gitlab-ce/issues/34659}}
\vspace{0.5mm}

Besides, during the classification of the 123 remaining issues, we identified and removed six issues that only request new features, bug corrections, or build failure fixes (\ie despite having a technical debt label, they are not related with TD). 
For example, we discarded an issue that reports:

\vspace{0.6mm}
\noindent{\em Commit count and other project statistics are incorrect.}\footnote{\url{https://gitlab.com/gitlab-org/gitlab-ce/issues/44726}}
\vspace{0.5mm}

After this step, we selected 117 SATD-I instances from GitLab.

\subsection{GitHub-based Projects}
\label{sec:github}

We also searched for SATD-I in open-source GitHub systems. 
We restricted the search to the top-5,000 most starred GitHub repositories, since 
stars is a commonly used proxy for the popularity of GitHub repositories~\cite{icsme2016, jss-2018-github-stars}. 
We used GitHub's REST API to search for all issues of such repositories that were closed in the period of December 15th, 2018 to June 15th, 2019---due to the same reasons explained for GitLab---and that include one of the following labels: {\em technical debt}, {\em Technical Debt}, and {\em debt}. 
We found 252 issues in 23 repositories. 
However, we decided to discard 34 issues from 19 repositories with  less than 10 issues. 
The rationale was to focus the study on repositories where labelling issues denoting TD is a common practice.

As for GitLab issues, the first author of this paper inspected all 218 initially selected issues (\ie 252 - 34 issues) and discarded 49 issues (22\%) that do not have a clear indication of representing an actual case of TD payment.
In the end, 169 SATD-I instances coming from four GitHub repositories were selected for inclusion in our dataset.

\subsection{Dataset Characterization}
\label{sec:characterization}

\begin{figure*}[ht!]
    \centering
    \subfloat[Time to close issues.]{
        \includegraphics[width=0.467\textwidth]{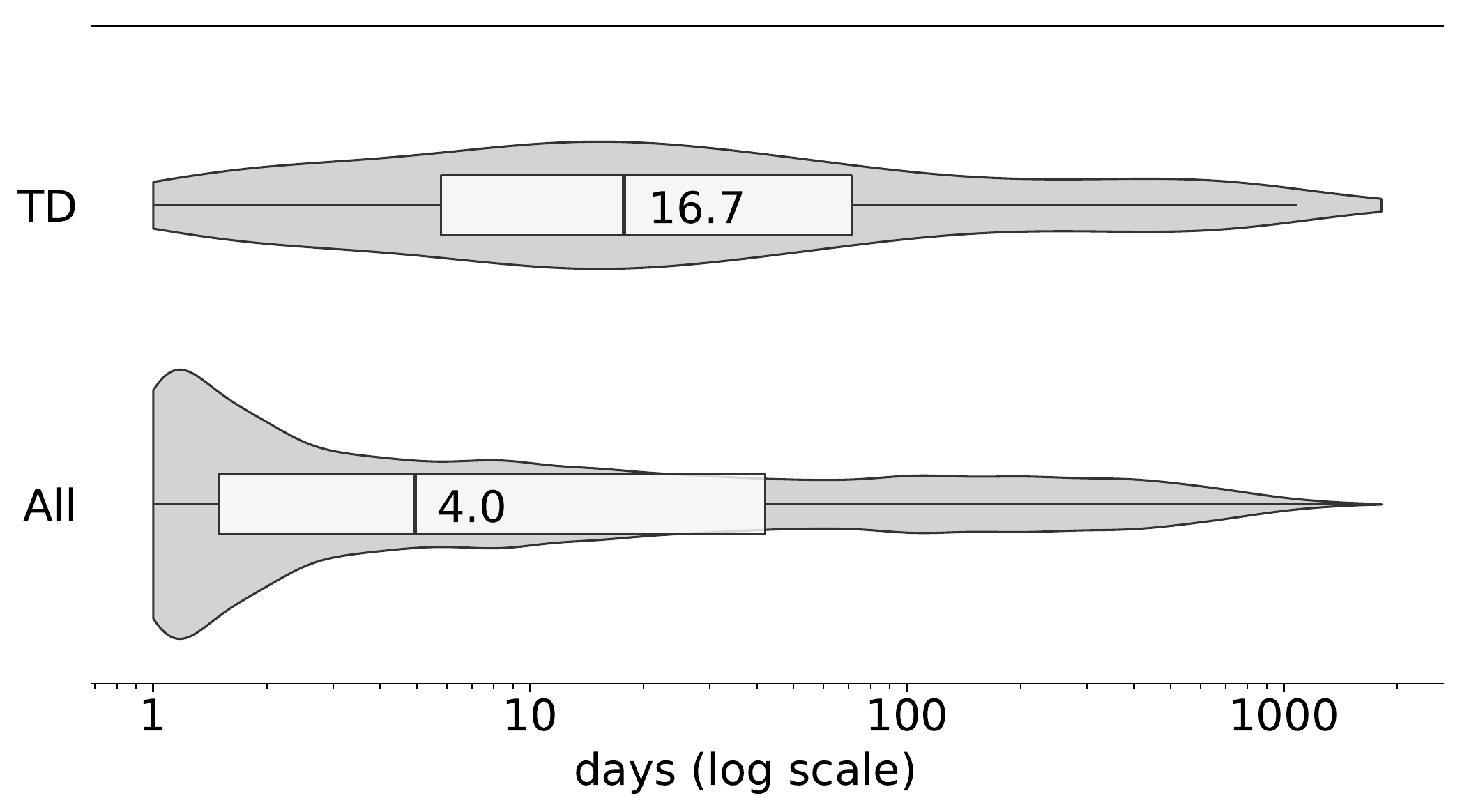}
        \label{fig:days_to_close}
    }
    \subfloat[Comments per issue.]{
        \includegraphics[width=0.467\textwidth]{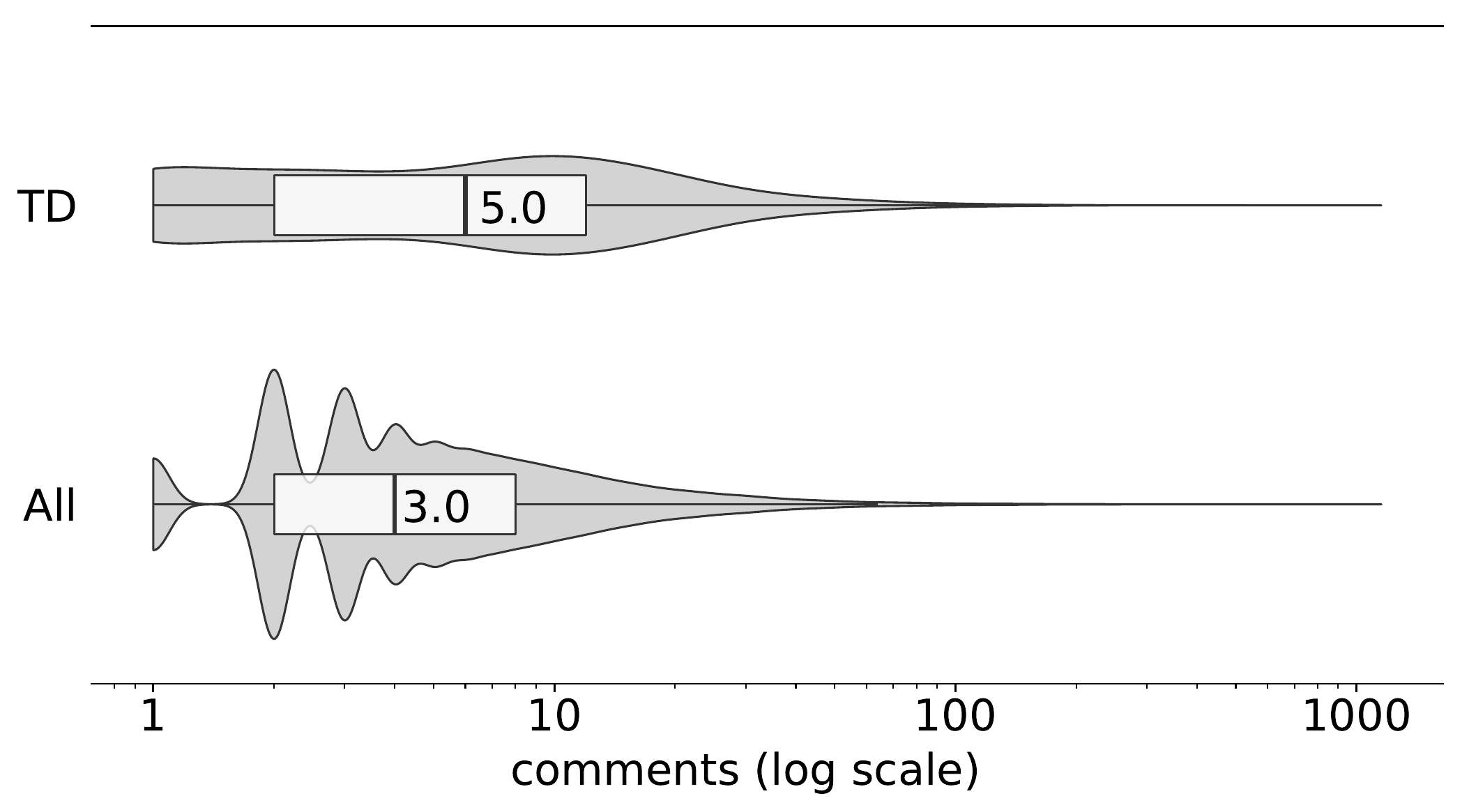}
        \label{fig:comments}
    }
    \hspace{0mm}
    \subfloat[Labels per issue.]{
        \includegraphics[width=0.467\textwidth]{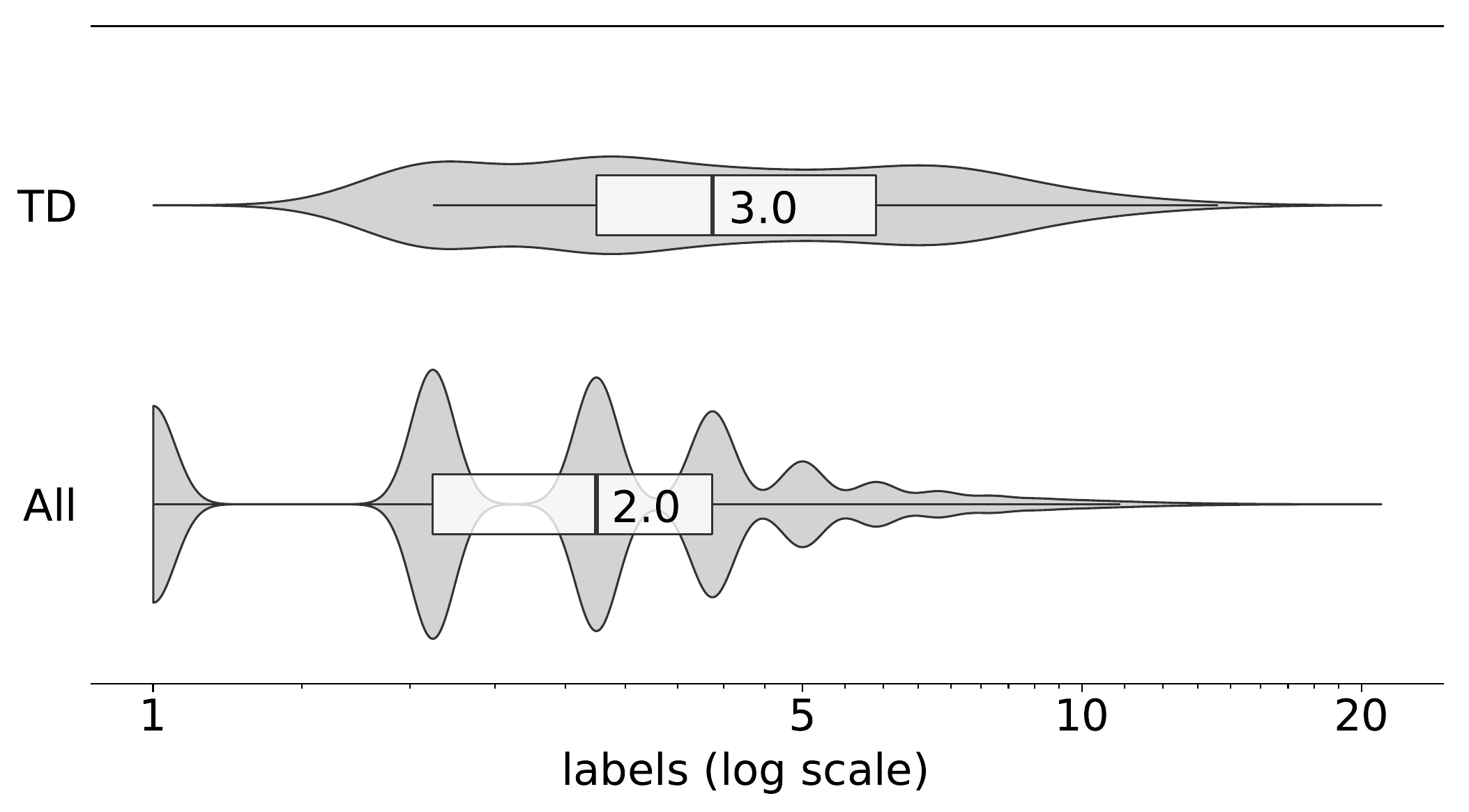}
        \label{fig:labels}
    }
    \subfloat[Code churn per issue.]{
        \includegraphics[width=0.467\textwidth]{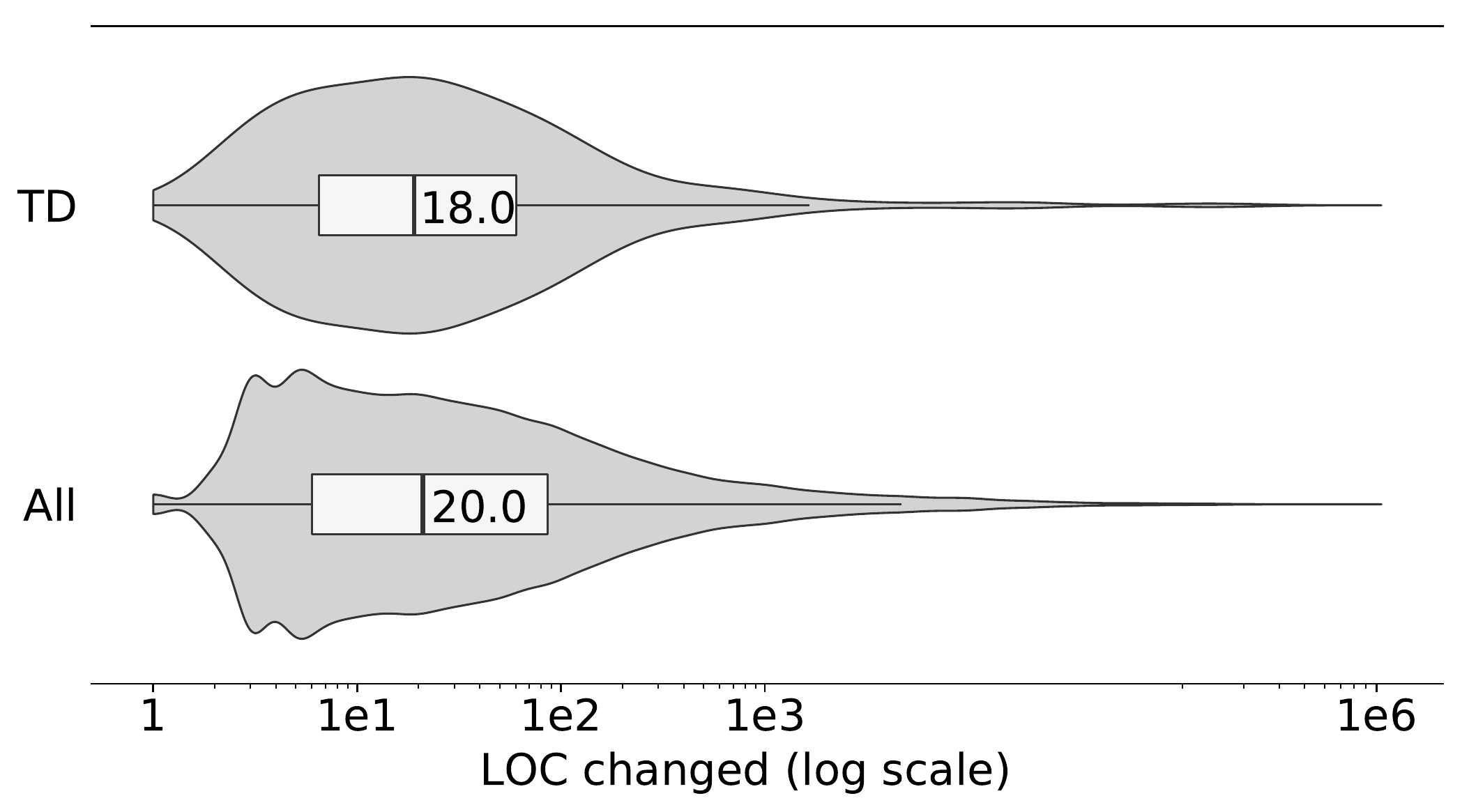}
        \label{fig:code_churn}
    }
    \caption{Distribution of days, comments, labels, and code churn per issue.}
    \label{fig:characterization}
\end{figure*}

Table I shows the name of the systems in our dataset, the tags they use to denote SATD-I and the number of issues selected in each system. As we can observe, there is
a concentration of issues
in GitLab-CE (40.9\%) and on \textsc{microsoft/vscode} (46.2\%), which is the popular IDE from Microsoft whose development history is now publicly available on GitHub. The remaining SATD-I instances come from \textsc{influxdata/influxdb} (7.3\%), \textsc{mirumee/saleor} (3.5\%), and \textsc{nextcloud/server} (2.1\%).
\textsc{influxdata/influxdb} is a framework for time series manipulation and visualization. \textsc{mirumee/saleor} is an open source eCommerce platform, and \textsc{nextcloud/server} is a framework for communicating with Nextcloud (a service for hosting files on the cloud).

\begin{table}[!h]
\vspace{-2mm}
	\centering
	\caption{Selected repositories }
	\label{table:issues-repos}
	\begin{tabular}{llrr}
		\toprule
		\textbf{Repository} & \textbf{Tag} & \textbf{SATD-I} & {\bf \%}\\ \midrule
		\textsc{microsoft/vscode} & debt & 132 & 46.2\%\\
		\textsc{gitlab/gilab-ce} & technical debt & 117 & 40.9\% \\
		\textsc{influxdata/influxdb} & Technical Debt & 21 & 7.3\%\\
		\textsc{mirumee/saleor} & technical debt & 10 & 3.5\%\\
		\textsc{nextcloud/server} & technical debt & 6 & 2.1\%\\ \midrule
		\multicolumn{2}{l}{\textbf{Total}} & 286 & 100\%\\
		\bottomrule
	\end{tabular}
\end{table}

Figure~\ref{fig:characterization} (in the next page) shows violin plots comparing the issues selected in the study with all other issues, i.e., the whole set of issues from the five studied systems. 
We can see that SATD-I takes more time to be closed (16.7 vs 4.0 days, median values).
They also have more comments (5 vs 3 comments) and labels (3 vs 2 labels).
These observations are statistically confirmed by applying the one-tailed variant of the Mann-Whitney U test (p-value $\leq 0.05$). 
Finally, the last chart (Figure~\ref{fig:code_churn}) shows the code churn of SATD-I versus all issues in our dataset.
The median code churn is 18 added/deleted lines (paid TD issues) versus 20 added/deleted lines (for all issues).
However, in this case, the distributions are not statistically different (p-value $= 0.13$).
\ie SATD issues are not different from other issues in terms of added and deleted lines of code.

\begin{formal}
Issues that pay technical debt trigger more discussions and take more time to be closed. 
However, paying SATD-I does not require larger code churns.
\end{formal}

\section{Overlap between SATD Types}
\label{sec:satd}

Self-admitted technical debt (SATD) refers to intentionally implemented Technical Debt, documented by developers either through code comments (SATD-C) or labelled issues (SATD-I), as we proposed in this paper.
However, it is not clear whether developers adopt only one of these approaches to admit their debts or there is an overlap between them.
In this section, we answer RQ1 by checking if SATD-I instances can be also classified as SATD-C. 


\subsection{Methodology}
\label{subsec:satad-method}

In order to analyze the 286 SATD-I instances to verify whether there are code comments admitting the same debt, we inspected the pull/merge-requests responsible for closing such issues (in GitLab, merge-requests are equivalent to pull-requests).
Specifically, we first collected the pull/merge-requests related to the issues (\ie the ones used to close the issue), retrieving the \textit{diff} of each associated commit.
Among the 286 analyzed issues, we found that 167 issues (58.4\%) were closed by an associated pull/merge-request.
The remaining issues were closed by internal contributors who directly committed the changes, for example.
Therefore, in this RQ we focus on these 167 issues, since in this case it is straightforward to navigate from the issues to the pull/merge-requests and then to the commits with the code that pays the SATD-I.

After this data collection phase, we analyzed all the \textit{deleted} lines from the commits paying the TD, in order to investigate whether there were comments indicating SATD-C. 
The rationale is that such comments should have been removed once the commits are responsible for closing the issues (thus, paying the debt).
We used SATDDetector~\cite{rw14} to automatically identify SATD-related comments in the deleted lines.
This tool applies NLP techniques to classify textual comments based on a pre-trained model.
It is based on a two-phase architecture, responsible for (i) building the model, and (ii) performing the classification.
To build the models, the tool's authors relied on a dataset with manually classified source code comments retrieved from eight open source projects~\cite{rw9}.
Finally, to perform the classification the tool applies a Naive Bayes Multinomial (NBM) technique.
\vspace{-2mm}
\subsection{Results}
\label{subsec:satd-results}

SATDDetector analyzed more than 4.7K \textit{diff} chunks associated to the 167 SATD-I instances.
The tool identified code comments indicating technical debt admission in 48 issues (28.7\%). 
Table~\ref{table:satd} details the results.
As we can see, the self-admitted comments are concentrated in two repositories: 
\textsc{microsoft/vscode}, with 26 TD-related issues where the TD was also self-admitted in the code (33.3\%); and
\textsc{gitlab/gilab-ce}, with 22 issues (21.7\%).
For the remaining repositories, no pull/merge-request was found to be analyzed.
Next, we provide examples for both repositories.
To facilitate the identification and discussion of the SATD-I instances in this paper, we label them using the initials of the repository name (\ie VS refers to \textsc{microsoft/vscde}; GL to \textsc{gitlab/gilab-ce}; IF to \textsc{influxdata/influxdb}; SL to \textsc{mirumee/saleor}; and NX to \textsc{nextcloud/ser-ver}).
The initials are then followed by an integer ID (\eg GL43 refers to issue 43 from GitLab).

\begin{table}[!h]
	\centering
	\caption{Overlap between SATD-I and SATD-C}
	\label{table:satd}
	\begin{tabular}{lrrr}
		\toprule
		\textbf{Repository} & \textbf{SATD-I} & \textbf{SATD-C} & {\bf \%}\\ \midrule
		\textsc{microsoft/vscode} & 78 & 26 & 33.3\%\\
		\textsc{gitlab/gilab-ce} &  89 & 22 & 21.7\% \\
		\textsc{influxdata/influxdb} & 0 & - & -\\
		\textsc{mirumee/saleor} & 0 & - & -\\
		\textsc{nextcloud/server} & 0 & - & -\\
		\midrule
		\textbf{Total} & 167 & 48 & 28.7\%\\
		\bottomrule
	\end{tabular}
\vspace{-3mm}	
\end{table}

\vspace{1mm}
\noindent\textbf{\textsc{microsoft/vscode}.} In the 26 occurrences of code-based self-admitted technical debt (33.3\%), developers include code comments to indicate temporary shortcuts or to guide contributors to fix the implemented solution.
As follows:

\vspace{2mm}

\noindent{\em HACK: This can be removed once this is fixed upstream xtermjs/ xterm.js\#1908} (VS122)

\vspace{2mm}

\noindent{\em If you want to provide a fix or improvement, please create a pull request against the original repository.} (VS27)

\vspace{2mm}



\noindent\textbf{\textsc{gitlab/gilab-ce}.} In 22 SATD-C instances (21.7\%), GitLab developers also indicate technical debt by adding code comments, as observed in these issues:

\vspace{2mm}

\noindent{\em TODO: remove eventHub hack after code splitting refactor} (GL90)

\vspace{2mm}

\noindent{\em Fixes or improvements to automated QA scenarios} (GL48)

\vspace{2mm}

\noindent{\em So you need to move all your global projects under groups or users manually before update or they will be automatically moved to the project owner namespace during the update. When a project is moved all its members will receive an email with instructions how to update their git remote URL. Please make sure you disable sending email when you do a test of the upgrade.} (GL84)

\vspace{2mm}

\begin{formal}
Only ~29\% of the issues that pay TD can be traced to SATD-C. In other words, 71\% of the studied issues document and pay TD that would not be possible to identify by considering only source code documentation.
\end{formal}

\section{SATD-I Classification}
\label{sec:class}

In this section, we present the results of our second research question, regarding the classification of the studied SATD-I.
We dedicate Section~\ref{subsec:class-method} to present the methodology applied to analyze the dataset.
Next, we present the classification results as follows:
first, in Section~\ref{subsec:td-class-results} we discuss issues related to {\sc Design} (the most popular type of TD) and its corresponding subclassification. 
In Section~\ref{subsec:other-class-results} we present the results for the other types of SATD-I instances.

\subsection{Methodology}
\label{subsec:class-method}

To identify the types of technical debt paid by developers, we carefully analyzed 286 SATD-I instances using {\em closed-card sort}~\cite{card-sorting}, a technique to classify a set of documents into predetermined categories. 
This technique involves the following steps: 
(i) defining the set of categories, 
(ii) initial reading of the issues,
(iii) classifying the issues by independent authors,
(iv) resolving conflicts.
We perform closed card sorting using categories previously elicited in the literature.
Specifically, we reused the categories described in a study performed by Li~\textit{et al.}~\cite{slr-td}.
In this work, the authors describe a systematic mapping to identify and analyze scientific papers on TD from 1992 to 2013. 
We classify the issues in our dataset according to the following categories proposed by Li~\textit{et al.}:

\begin{itemize}
    \item {\sc Design}: refers to technical shortcuts used in internal method design and high-level architecture.
    \item {\sc UI}: refers to debt on the elements of user interfaces.
    \item {\sc Tests}: refers to the absence of tests or to workarounds on existing code for testing.
    \item {\sc Performance}: refers to debt that affects system performance (\eg in time and memory usage).
    \item {\sc Infrastructure}: refers to debt on third-party tools, obsolete technologies or deprecated APIs.
    \item {\sc Documentation}: refers to insufficient, incomplete, or outdated documentation.
    \item {\sc Code Style}: refers to code style violations.
    \item {\sc Build}: refers to debt in the build system, as when using scripts that make the build more complex or slow.
    \item {\sc Security}: refers to shortcuts that expose system data or compromises user permission access.
    \item {\sc Requirements}: refers to debt on requirements specification that leads to implementation problems.
\end{itemize}

Since {\sc Design} was the most popular case of SATD-I (as we found in our first round of classification), we decided to perform a sub-classification of this type of issues.
Thus, we defined four categories: 

\begin{itemize}
    \item {\sc Complex Code}: refers to intra-method poorly implemented code.
    \item {\sc Architecture}: refers to high-level design problems, including inadequate organization of packages.
    \item {\sc Clean Up}: refers to the elimination of obsolete or dead code.
    \item {\sc Code Duplication}: refers to code clones that should be removed to improve maintainability.
\end{itemize}

After defining the mentioned categories, three authors of this paper manually analyzed the issues, by reading their descriptions and existing discussions, with the goal of assigning one (or more) categories. 
Each issue was analyzed by two independent authors.
In 178 cases (62.2\%) they agreed in the first proposed classification.
For the {\sc Design} subclassification (169 issues), the authors agreed in 92 cases (54.4\%).
In a final step, the authors discussed each conflict and reached a consensual classification.
We use the same identification of Section~\ref{sec:satd} to label the following discussion.

\subsection{SATD-I related to Design}
\label{subsec:td-class-results}

With 169 occurrences (59.1\%), most of the selected issues refer to {\sc Design} debt.
In this case, we classified {\sc Design} SATD-I into four subcategories, as presented in Table~\ref{tab:design-classification}.
As we can see, {\sc Complex Code} is the type of {\sc Design} TD more commonly paid by developers (43.8\%), followed by {\sc Architecture} (33.7\%), {\sc Clean Up} (18.9\%), and {\sc Code Duplication} (3.5\%).
Next, we describe and provide examples for each {\sc Design} category.

\begin{table}[!ht]
\vspace{-3mm}
\centering
\renewcommand{\arraystretch}{1.3}
\caption{Design SATD-I Classification}
\label{tab:design-classification}
\footnotesize 
\begin{tabular}{lrr} \toprule
\textbf{Technical Debt} & \textbf{Occ.} & \textbf{\%}\\ \midrule
\textsc{Complex Code} & 74 & 43.8\% \\
\textsc{Architecture} & 57 & 33.7\%\\
\textsc{Clean Up} & 32 & 18.9\%\\
\textsc{Code Duplication} & 6 & 3.5\%\\
\bottomrule \end{tabular}

\vspace{-3mm}
\end{table}

\noindent\textbf{Complex Code.} In 74 cases (43.8\%), {\sc Design} issues are related to technical shortcuts that developers take when implementing methods.
In this case, the payment involves changes only in the single method where the debt is located. 
As an example, the following issues are related to this type of {\sc Design} SATD-I:

\vspace{2mm}

\noindent{\em We should unify naming related to checkout functionality, as currently, we're mixing ``checkout'' with ``cart'', which leads to confusion when reading the code. I recommend that we settle on the name ``checkout'' and rename the Cart model and all other occurrences of cart.} (SL1) 

\vspace{2mm}

\noindent{\em Currently, errors is an optional list of optional errors. While returning an empty list is probably not needed, current type forces the client to make sure the errors themselves are not null.} (SL10) 

\vspace{2mm}

\noindent\textbf{Architecture.} With 57 occurrences (33.7\%), the second type of {\sc Design} TD most commonly paid by developers is related to high-level design flaws.
To pay this type of debt, it is usually necessary to make changes in the organization of packages and modules, for example.
The following issues are related to this type of SATD-I:

\vspace{2mm}

\noindent{\em This class is way too big for its own good. For example, there's no need for it to update a project's main language in the same job/thread/process as the other work.} (GL43) 

\vspace{2mm}

\noindent{\em The root of the TimeMachine tree contains a TimeSeries component. This component handles fetching time series data used in the TimeMachine (\ldots) The aim of this refactor would be to move all state from the TimeSeries component into Redux and all logic into a thunk.} (IF12) 

\vspace{2mm}

\noindent\textbf{Clean Up.} Next, issues related to the presence of obsolete or dead code represents the third most common type of SATD-I, with 32 instances (18.9\%).
As an example, the following issue is related to this type of {\sc Design} TD:

\vspace{2mm}

\noindent{\em In Milestone 11.4, we introduced personal\_access\_tokens. token\_digest, so we can now remove personal\_access \_tokens.token.} (GL47)


\vspace{2mm}



\noindent\textbf{Code Duplication.} Finally, with 6 occurrences (3.5\%), the least common type of {\sc Design} SATD-I refers to duplicated code, as illustrated by the following issue:

\vspace{2mm}

\noindent{\em There has been a lot of duplication of frontend code between Protected Branches and Protected Tag feature, this issue is intended to reduce duplication.} (GL81) 



\subsection{Other Types of SATD-I}
\label{subsec:other-class-results}

Table~\ref{tab:td-classification} presents the classification of the remaining SATD issues.
As we can see, if we do not count issues related to {\sc Design} (59.1\%), the most common type of paid TD refers to {\sc UI} (10.1\%), {\sc Tests} (8.7\%), and {\sc Performance} (8\%).
Next, we describe these categories.

\begin{table}[!h]
\centering
\renewcommand{\arraystretch}{1.3}
\caption{Other Types of SATD-I}
\label{tab:td-classification}
\footnotesize 
\begin{tabular}{lrr} \toprule
\textbf{Technical Debt} & \textbf{Occ.} & \textbf{\%}\\ \midrule
\textsc{UI} & 29 & 10.1\% \\
\textsc{Tests} & 25 & 8.7\% \\
\textsc{Performance} & 23 & 8\% \\
\textsc{Infrastructure} & 18 & 6.3\% \\
\textsc{Documentation} & 12 & 4.2\% \\
\textsc{Code Style} & 8 & 2.8\% \\
\textsc{Build} & 4 & 1.4\% \\
\textsc{Security} & 3 & 1.1\% \\
\textsc{Requirements} & 3 & 1.1\% \\
\bottomrule \end{tabular}

\vspace{-3mm}
\end{table}

\vspace{2mm}
\noindent\textbf{UI.} With 29 occurrences (10.1\%), the second most common type of SATD-related issues refers to debt on user interface code.
In this case, developers implemented shortcuts that result in usability flaws, as mentioned in the following issue:





\vspace{2mm}

\noindent{\em Today there is a ``Building...'' label appearing around the problems entry when building a project. I think this originates from a time where we did not have support to show progress in the status bar.} (VS29)


\noindent\textbf{Tests.} In 25 cases (8.7\%), SATD-related issues report the absence of tests or request improvements on existing tests.
The following issue illustrates this type of TD:



\vspace{2mm}

\noindent{\em I want to have some tests that will give me a better perspective for usage of DB queries under GraphQL API. I would like to have explicit logic to validate that it works as expected.} (SL3)

\vspace{2mm}

\noindent\textbf{Performance.} With 23 occurrences (8\%), the fourth most common type of SATD-I is related to performance concerns, in terms of time or memory usage.
This is illustrated as follows:

\vspace{2mm}

\noindent{\em underscore.js is bundled in vendor/core.js but it's the unminified version. Can we replace it with the minified version? The file size is a lot smaller.} (NX1)

\vspace{2mm}

\noindent{\em Every widget and actions in each extension has a global listener to check if there is a change and update itself. This causes 100s of listeners being added to a global event.} (VS65)










\begin{formal}
SATD-I is paid mostly to fix {\sc Design} flaws ($\sim$60\%).
But we also found paid TD related to {\sc UI} (10\%), {\sc Tests} (9\%), and {\sc Performance} (8\%), for example.
\end{formal}

\section{Survey with Developers}
\label{sec:quali}

To answer the remaining research questions, we perform a survey with the developers responsible for closing the 286 SATD-I instances studied in this paper.
In this section, we first present the methodology followed in this survey (Section~\ref{subsec:survey-method}). After that, we present the results for each research question (Sections~\ref{subsec:td-origin} and~\ref{subsec:td-payment}).

\subsection{Methodology}
\label{subsec:survey-method}

We conduct a survey to reveal the reasons why developers introduce technical debt in their code, the motivations of SATD-I payment, and the maintenance problems associated to TD.
For that, we sent emails to developers that closed the SATD issues studied in this paper. 
Specifically, we selected from our dataset developers with public email address who were responsible for (i) closing a specific issue; or (ii) accepting a pull/merge-request that closes the issue.
From the total of 286 issues, we retrieved a list of 85 distinct emails. 
In the cases where the same developer was responsible for more than one issue, we selected the most recently closed one. 

For each developer, we sent the questionnaire between July 11th and August 14th, 2019 (which represents an interval of at most six months after the date the issues were closed).
Figure~\ref{fig:email} shows the template of the survey email.
First, we presented the issue that represents the debt paid by the developer.
Next, we proposed three questions with the goal of
(1) investigating the reasons why developers pay technical debt; 
(2) unveiling maintenance problems caused by TD; and 
(3) understanding the intentions behind TD insertion.
Questions (1) and (2) were open-ended, while question (3) provided two predefined options, reused from the technical debt quadrant proposed by Martin Fowler~\cite{fowler-quadrant}.
Although developers could simply select one of the answers, we allowed them to provide their own answers or to include comments to predefined answers.

\begin{figure}[t!]
\centering
\noindent\mbox{
	\begin{minipage}{0.44\textwidth}\footnotesize{{\it 
	    \hrule height 0.03cm 
    	\vspace{2mm}
        I figured out that you closed the following issue from [repository name]:\\
        
        [issue title]
        [issue link]\\
        
        which is labeled as [TD-related tag].\\
        
        I kindly ask you to answer the following questions:\\
        
        1. Why did you decide to pay this TD?\\
        
        2. Could you describe the maintenance problems caused by this TD?\\
        
        3. Could you classify this TD under the following categories:\\
        
            a. It was deliberately introduced to ship earlier\\
            b. When it was introduced, we were not aware about the best design\\
            c. Other answers (please clarify)
        \vspace{2mm}
        \hrule height 0.03cm 
    	}}		
\end{minipage}}
\caption{Email sent to developers who paid SATD-I.}
\label{fig:email}
\vspace{-3mm}
\end{figure}

We received 30 answers coming from developers of four repositories (\ie response rate of 35.3\%). 
Table~\ref{table:survey-answers} details the number of emails sent and the answers received per repository.
{\sc microsoft/ vscode} and {\sc influxdata/influxdb} have the highest response rate (both with 40\%). 
However, they do not represent the majority of the answers, once we received 23 answers from GitLab developers.

\begin{table}[!h]
    \vspace{-2mm}
	\centering
	\caption{Survey answers}
	\label{table:survey-answers}
	\begin{tabular}{lrrr}
		\toprule
		\textbf{Repository} & \textbf{Sent} & \textbf{Answers} & {\bf \%}\\ \midrule
		\textsc{microsoft/vscode} & 10 & 4 & 40\%\\
		\textsc{influxdata/influxdb} & 5 & 2 & 40\%\\
		\textsc{gitlab/gilab-ce} & 66 & 23 & 34.9\% \\
		\textsc{nextcloud/server} & 3 & 1 & 33.3\%\\ 
		\textsc{mirumee/saleor} & 1 & 0 & 0\%\\\midrule
		\textbf{Total} & 85 & 30 & 35.3\%\\
		\bottomrule
	\end{tabular}
	\vspace{-2mm}
\end{table}

To interpret the survey answers (1) and (2), the first author followed an {\em open card-sorting}~\cite{card-sorting}.
This technique is used to identify {\em themes} (\ie patterns) in textual documents through the following steps:
(i) identifying themes from the answers, 
(ii) reviewing the themes to find opportunities for merging, and 
(iii) defining and naming the final themes. 
During the analysis, one answer was discarded because the developer did not actually discussed the issue.
In a final step, the last author reviewed and confirmed the proposed themes.
In the following discussion, we label the quotes with D1 to D29 to indicate developers answers.

\subsection{Why do developers introduce SATD-I?}
\label{subsec:td-origin}

\begin{figure}[!h]
\centerline{\includegraphics[width=.475\textwidth]{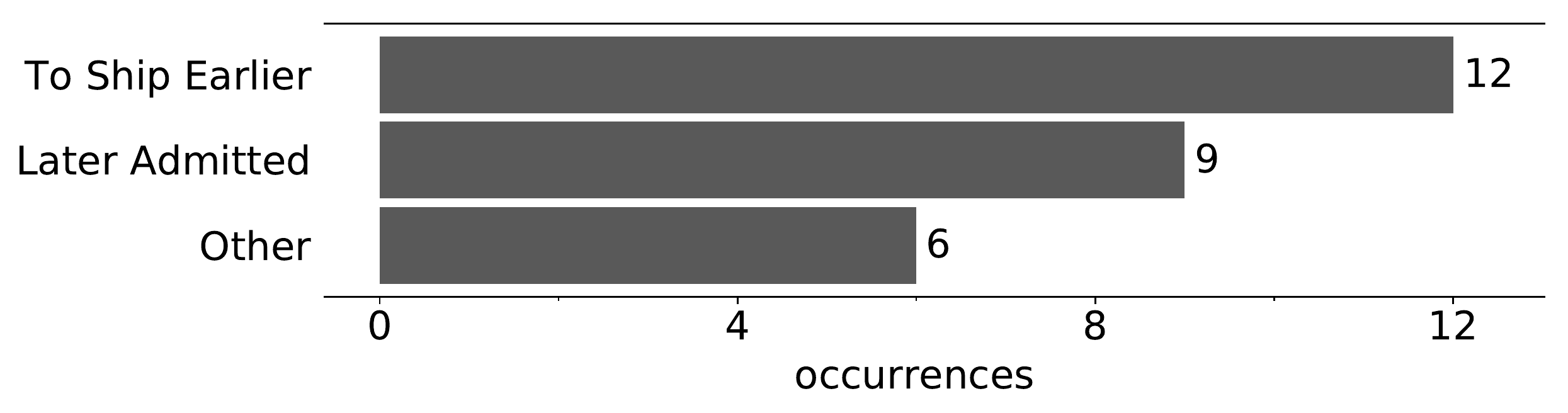}}
\caption{Reasons for introducing SATD-I.}
\label{fig:td-introduction}
\vspace{-3mm}
\end{figure}

To answer this question, we provided two predefined options: the first is related to developers decision of  introducing TD as a choice for agility.
The second corresponds to the scenario where developers only perceived the debt after it was introduced.
We also left the opportunity for developers to clarify their answers and provide further information.
Figure~\ref{fig:td-introduction} presents the obtained results.
As we can observe, most of the studied debts were introduced by developers to ship earlier (12 answers). 
In nine cases, developers were not aware of the TD when it was introduced.
Finally, six developers provided other motivations.
Next, we detail each of these reasons and provide quotes from extra comments discussed by developers.

\vspace{2mm}
\noindent\textbf{SATD introduced to ship earlier.} In 12 answers (44.5\%), developers confirmed that the technical debt was introduced to speed up development.
In other words, to deliver faster, developers consciously added shortcuts in their code which were expected to be fixed in the future. To remind about this fact, they also decided to document the TD using an issue.
D7, D16, and D9 provided further details for this reason:

\vspace{2mm}

\noindent{\em It was thoroughly discussed and weighed up before we take the decision to accept the TD to be dealt with on a next release. The TD wasn't introducing any critical performance issues or bugs to the system. Furthermore, we were confident that we could fix the TD in the next release, which happened.} (D7)

\vspace{2mm}

\noindent{\em I think that usually when we introduce a technical debt it either helps us to ship something earlier/faster or makes first iteration of implementation much easier in general.} (D16)

\vspace{2mm}

\noindent{\em We were aware and were ok with the implementation for now as long as we fixed it afterwards} (D9)

\vspace{2mm}

\noindent\textbf{SATD admitted after introduction.} For nine developers (33.3\%), the debt was originated by their lack of understanding about the best design solution at the time the code was initially implemented. After discovering or facing the TD, they decided to admit it opening an issue. The following answers illustrate this scenario:



\vspace{2mm}

\noindent{\em  We figured we'd never hit ``that" usecase. But we did.} (D23)

\vspace{2mm}

\noindent{\em The class just grew over time without planning.} (D15)

\vspace{2mm}

\noindent\textbf{Other reasons.} Finally, six developers provided other reasons for introducing TD in their code (22.2\%).
Answers include the advent of new technologies that turned the old code a debt, and also the mischoice of design alternatives.
This is illustrated in the following examples:

\vspace{2mm}

\noindent{\em It slowly became TD, while at the time of the initial development it was most likely fine to code that way.} (D29)

\vspace{2mm}

\noindent{\em I think the original author just overlooked that exposing these methods wasn't really needed.} (D13)

\begin{formal}
In most of the cases, SATD-I is introduced as a deliberate choice for agility (44.5\%).
\end{formal}

\subsection{Why do developers pay SATD-I?}
\label{subsec:td-payment}

In order to investigate the reasons why developers pay SATD-I, we combine answers from questions (1) and (2) of our e-mails (Section~\ref{subsec:survey-method}).
First, we directly asked developers the reasons that drive such payment. 
Next, we complement our findings by eliciting a list of associated maintenance problems.

We first identified five distinct reasons why developers pay SATD-I, as reported in Table~\ref{table:motivations-td}.
As we can see, reducing TD interest is the most common motivation for SATD-I payment (65.5\%), followed by the desire to have a clean code (27.6\%).
In some cases, a given answer produced more than one motivation.
This explains why the number of occurrences is higher than the number of answers (29 answers).
Next, we discuss these reasons.

\begin{table}[!h]
\vspace{-2mm}
	\centering
	\caption{Reasons why developers pay SATD-I}
	\label{table:motivations-td}
	\begin{tabular}{lrr}
		\toprule
		\textbf{Why did you decide to pay this TD?} & \textbf{Occ.} & \textbf{\%} \\ \midrule
		\textit{To reduce TD interest} &  19 & 65.5\%\\
		\textit{To clean code} &  8 & 27.6\%\\
		\textit{To get familiarised with the codebase} &  2 & 6.9\%\\
		\textit{To collocate with other related work} &  2 & 6.9\%\\
		\textit{To increase test coverage} &  1 & 3.5\%\\
        \bottomrule
	\end{tabular}
\vspace{-3mm}
\end{table}

\vspace{2mm}

\noindent{\bf To reduce TD interest.} With 19 answers (65.5\%), the most common reason for paying SATD-I is to reduce TD interest.
Although developers did not directly mention the term \textit{interest}, eliminating the maintenance burden caused by the studied issues was mentioned in several answers.
For example, D1 and D19 mention this motivation:






\vspace{2mm}

\noindent{\em This was adding extra maintenance for me.} (D1)

\vspace{2mm}

\noindent{\em The component was growing too big, making it difficult to maintain.} (D19)

\vspace{2mm}
\noindent{\bf To clean code.} In eight cases (27.6\%), technical debt payment is related to the desire of having a clean codebase (\eg to reduce code complexity and remove duplication).
For example, the following answers are related to this motivation:

\vspace{2mm}

\noindent{\em To keep the code clean and easy to read/maintain.} (D27)

\vspace{2mm}

\noindent{\em To get the benefits of a cleaner code (\ldots). After fixing the TD, understanding the code got easier. It also got smaller.} (D7)

\vspace{2mm}













\begin{formal}
Technical debt is periodically paid to reduce its interests (66\%), and to clean code (28\%).
\end{formal}


We also asked the survey participants to comment on the specific maintenance problems that motivated them to close the studied SATD-I instances.
Table~\ref{table:maintenance-problems} presents the list of the most common answers.
In this case, three answers (out of the 29 analyzed) were discarded because they were not clear.
According to the remaining answers, TD is mostly responsible for slowing down code evolution, increasing maintenance effort due to duplicated code, and making it harder to read and understand code. 
The three problems occurs with the same frequency (six answers for each).

\begin{table}[t]
	\centering
	\caption{Maintenance problems caused by technical debt}
	\label{table:maintenance-problems}
	\begin{tabular}{lrr}
		\toprule
		\textbf{What problems are caused by this TD?} & \textbf{Occ.} & \textbf{\%} \\ \midrule
		\textit{Code was difficult to evolve} &  6 & 20.7\%\\
		\textit{Duplicated code was demanding extra effort} &  6 & 20.7\%\\
		\textit{Code was difficult to read and understand} &  6 & 20.7\%\\
		\textit{Code performance was poor} &  5 & 17.2\%\\
		\textit{Code was error-prone} &  4 & 13.8\%\\
		\textit{UI presented visual defects } &  1 & 3.5\%\\
        \bottomrule
	\end{tabular}
\vspace{-3mm}
\end{table}

\begin{formal}
Technical debt is commonly responsible for slowing down code evolution, duplicating maintenance effort, and making it harder to read and understand code.
\end{formal}

\section{Implications}
\label{sec:discussion}

This section presents the study implications on tool support and process improvement.

\vspace{1.5mm}
\noindent\textbf{Tool Support.}
In RQ1, we show that only 29\% of the SATD-I studied instances are also admitted through code comments.
Instead, developers tend to create issues reporting debts, labelling them with TD-related tags.
Besides, in RQ3 developers point that the majority of SATD-I instances are intentionally created to ship earlier.
Therefore, these results reinforce that developers usually decide to follow the ``done is better than perfect'' maxim, implementing suboptimal code solutions in order to deliver on time.
To tackle this problem, we envision research on new tools to allow developers to explicitly label new debts inserted on GitHub/GitLab-based systems.
Specifically, such tools would be responsible for asking pull/merge-requests authors whether the contribution contains any type of TD.
In the cases when they admit it, the tool would suggest the automatic opening of a follow-up issue, tagged with a SATD-related label (as in the issues we studied in this paper).
These tools would be effective for various roles of contributors, since:
(i) core-developers would benefit from managing code quality and better reviewing contributions,
(ii) pull/merge-request authors would feel responsible for their own debts (and possibly will come back to pay them), and
(iii) newcomers could pay these issues as a way to get familiarized with the code (as we will better discuss in the next implication).
Moreover, the number of open TD-related issues could be used as a metric to measure the quality of the system.
Although there are several approaches to automatically identify TD on source code (\eg~\cite{rw3, rw12, rw14}), we claim this tool is based on developers feedback right after TD insertion. 
It would also be independent of programming language. 

\vspace{1.5mm}
\noindent\textbf{Process Improvement.} Among the reasons that drive developers to pay technical debt (elicited in RQ4), we identified two motivations explicitly related to software development process:
\textit{to reduce TD interest} (with 65.5\% of occurrences) and 
\textit{to get familiarized with the codebase} (with 7\%).
In other words, this result shows that paying technical debt represents an actual activity introduced in the development process of the studied projects to preserve internal quality and to train new contributors on the structure of the code.
Therefore, we extrapolate this finding by suggesting two particular implications:
first, we suggest the formalization of technical debt payment as an actual activity on modern development processes.
In agile processes these activities could be introduced as {\em slacks} to keep developers productive, for example.
Second, we advocate that paying technical debt may be included on onboarding activities for new team members~\cite{Steinmacher:newcomers}.
In this case, team leaders would {\em ``delegate this work to newcomers to give them easy stuff to familiarize themselves with the work process''} (D19).


\section{Threats to Validity}
\label{sec:threats}

\noindent\textbf{External Validity.} This study is restricted to 286 closed issues classified as technical debt according to SATD-related labels.
Although the issues were selected from relevant repositories, maintained by organizations like Microsoft and GitLab, we cannot generalize our findings to other systems, especially to the ones that apply different approaches to manage technical debt (\ie do not use TD-related labels).
Moreover, the results discussed in Section~\ref{sec:quali} are based on the opinion of 29 developers, mostly from GitLab.
Despite that, we claim that the obtained response rate (35.3\%) represents a relevant mark in typical software engineering studies, with valuable insights from developers actually responsible for paying SATD-I.

\vspace{1.5mm}
\noindent\textbf{Internal Validity.}
First, we selected the studied issues by using TD-related labels as a proxy for technical debt identification.
However, as discussed by Kruchten~\textit{et al.}~\cite{td-definitions}, the concept of technical debt has been diluted since its original proposition.
Thus, the misunderstanding of this concept by those who classified the TD-labeled issues would affect the results of our study.
To alleviate this threat, the first author of this paper carefully analyzed the initial dataset of 406 TD-labeled issues, and discarded 120 issues (29.6\%) that did not have a clear indication of TD payment.
Second, we should mention the subjective nature of both closed- and open-card sort used in Sections~\ref{sec:class} and~\ref{sec:quali}. 
Despite the rigor followed by the authors to perform these classifications, the replication of this activity may lead to different results.
To mitigate this threat, special attention was paid during the discussions to resolve conflicts and to assign the final themes.
Third, against our belief, the correctness of developers answers is also a threat to be reported.
To alleviate it, we restricted our study to issues closed in the last six months, which was important to guarantee a higher response rate and to increase answers reliability.
Finally, the classification of SATD-C in RQ1 relies on the implementation of a third-party tool proposed in previous studies.
Although the tool is vastly used to this end and represent the state-of-the-art to detect SATD-C, the possibility of false positives in the classification may also be reported as a threat. 

\vspace{1.5mm}
\noindent\textbf{Construct Validity.}
This threat concerns the selection of the TD-related issues on GitHub-based projects.
As discussed in Section~\ref{sec:data}, besides GitLab's issues (that we had previous knowledge about its labeling practices), in order to search for analogous issues on GitHub, we mined repositories that included ``technical debt'', ``Technical Debt'', and ``debt'' as labels.
Therefore, it is possible that other tags are used to denote TD-related issues.

\section{Related Work}
\label{sec:related-work}
 
Recent research on technical debt derive from the concept of Self-Admitted TD (SATD) presented by Potdar and Shihab~\cite{potdar2014}. 
In this work, the authors observe that developers commonly document TD through source code comments.
Through the analysis of more than 100K code comments, they find that (i) 2.4\%--31\% of source code files contain self-admitted technical debt, (ii) experienced developers tend to introduce more debts, and (iii) 26\%--63\% of SATD gets removed.
Bavota and Russo~\cite{rw5} replicate this study on a larger dataset that includes 600K commits and 2 billion comments.
The authors first confirm the previous findings, observing that the amount of SATD increases over time and tend to survive a long time in the system.
In this paper, we extend their findings by observing that developers may acknowledge TD out of source code. In this case, we define traditional code-base SATD as SATD-C, and investigate the occurrence of issue-based SATD, i.e.,~SATD-I.

Maldonado~\textit{et al.}~\cite{rw20} investigate five Java open source projects with the purpose of examining the amount of TD removed, who performs the removal, how long it lives in a project, and what activities lead to the removal.
As a result, the authors show that the majority of SATD is removed from projects by the same developer who introduced the debt (\ie self-removed), as part of bug fixing activities and the addition of new features.
Zampetti~\textit{et al.}~\cite{rw2} perform a follow-up study, based on the dataset elicited by Maldonado~\textit{et al.}~\cite{rw20}, to quanti- and qualitatively investigate how self-admitted technical debt is removed. 
Specifically, the authors assess the amount of SATD removals that are actually accidental transformations, as well as the extent to which SATD removals are documented in commit messages. 
Although these studies also investigate TD removal, we claim that their findings are restricted to SATD-C instances.



Sierra~\textit{et al.}~\cite{rw1} investigate the possibility of using source code comments that indicate technical debt to resolve architectural divergences. 
The authors used a dataset of previously classified SATD comments to trace architectural divergences in an open-source system. 
They found that 14\% of divergences could be directly traced.
Therefore, they stand that it is viable to use SATD comments as an indicator of architectural divergences.
Farias~\textit{et al.}~\cite{id-br}, Huang~\textit{et al.}~\cite{rw12} and Liu~\textit{et al.}~\cite{rw14} also identify SATD by mining source code comments.
Moreover, other studies propose the use of natural language processing (NLP) techniques to support SATD identification~\cite{rw9}. 
For example, Flisar and Podgorelec~\cite{rw11}, Huang~\textit{et al.}~\cite{rw12}, Ren~\textit{et al.}~\cite{rw13}, and Fahid~\textit{et al.}~\cite{rw17} use machine learning techniques for automating SATD detection. 
Dai and Kruchten~\textit{et al.}~\cite{rw15} improves these approaches by identifying non-code-level technical debt.
Maldonado \textit{et al.}~\cite{rw3} proposed a technique to precisely identify SATD, outperforming the current state-of-the-art, based on fixed keywords and phrases. 
The proposed technique achieved good accuracy for between 80\%--90\% of the cases.
We take advantage of SATDDetector to identify SATD-C in RQ1.

\section{Conclusion}
\label{sec:conclusion}

In this paper, we performed three complementary studies with the purpose of answering four research questions on Self-Admitted Technical Debt documented through labelled issues (SATD-I).
We analyzed 286 SATD-I to
(1) investigate the overlap between code-based SATD (SATD-C) and issue-based SATD (SATD-I), 
(2) identify the types of SATD-I more frequently paid,
(3) understand the intentions behind SATD-I insertion, and
(4) investigate the reasons why developers pay SATD-I.
As a result, we showed that SATD-I instances take more time to be closed, although they are not more complex in terms of code churn. 
We also revealed that only 29\% of the studied SATD-I instances can be tracked to source code comments, and that 45\% of the studied debt was introduced to ship earlier. 
In almost 60\% of the cases, we found that SATD-I is related to Design flaws (with a concentration on method-level debt in 44\% of this total). 
Moreover, our results indicated that most developers paid SATD-I to reduce its interest, and to have a clean code.
As practical implications of our studies, we suggest novel tools to support technical debt management. 
We also discuss how SATD-I payment activities can be introduced as part of software development processes.

As future work, we first intend to enlarge our dataset of SATD-I by mining other tags that may denote TD-related issues. 
After that, we envision an in-depth analysis of the code transformations performed to pay these debts.
Based on this dataset of transformations, we may develop tools and techniques to guide developers on TD payment (\eg by recommending how to perform changes that contribute to the actual removal of the debt).
As a final note, our data is publicly available at: \url{http://doi.org/10.5281/zenodo.3701471}.

\section*{Acknowledgments}
\noindent We thank the 30 developers who participated in
our survey and shared their ideas and practices about technical debt payment. 
This research is supported by grants from FAPEMIG, CNPq, and CAPES.

\balance

\bibliographystyle{ACM-Reference-Format}
\bibliography{bib}

\end{document}